\documentclass[journal,10pt]{IEEEtran}

\usepackage[T1]{fontenc}

\usepackage{graphicx,amssymb,amsmath,amsthm,verbatim}
\usepackage[cmintegrals]{newtxmath}
\usepackage{mathtools}
\usepackage{subfigure}
\usepackage{setspace}
\usepackage{setspace}
\usepackage{multirow}
\usepackage{pifont}

\usepackage[outdir=./]{epstopdf}
\usepackage{soul}
\usepackage{algorithm}
\usepackage{algorithmic}
\usepackage{xcolor}
\usepackage{cases}
\usepackage{cite} 

\ifCLASSINFOpdf
  
\else
  
\fi

\usepackage{amsmath}

\begin{document}

\title{Average AoI Minimization for Energy Harvesting Relay-aided Status Update Network Using Deep Reinforcement Learning}

\author{Sin-Yu Huang and Kuang-Hao (Stanley) Liu,~\IEEEmembership{Member,~IEEE}
\thanks{S-Y.~Huang is with the Institute of Computer and Communication Engineering,
National Cheng Kung University, Tainan, Taiwan 701. E-mail: {\tt q36104030@gs.ncku.edu.tw}}
\thanks{K-H.~Liu is with the Institute of Communications Engineering,
National Tsing Hua University, Hsinchu, Taiwan 300044. E-mail: {\tt khliu@ee.nthu.edu.tw}}
}

\maketitle

\begin{abstract}
A dual-hop status update system aided by energy-harvesting (EH) relays with finite data and energy buffers is studied in this work. To achieve timely status updates, the best relays should be selected to minimize the average age of information (AoI), which is a recently proposed metric to evaluate information freshness. The average AoI minimization can be formulated as a Markov decision process (MDP), but the state space for capturing channel and buffer evolution grows exponentially with the number of relays, leading to high solution complexity. We propose a relay selection (RS) scheme based on deep reinforcement learning (DRL) according to the instantaneous channel packet freshness and buffer information of each relay. Simulation results show a significant improvement of the proposed DRL-based RS scheme over state-of-art approaches.  
\end{abstract}


\IEEEpeerreviewmaketitle

\setlength{\abovedisplayskip}{4.0pt} 
\setlength{\belowdisplayskip}{3.0pt}
\setlength{\parskip}{-1pt}

\section{Introduction}
Emerging Internet of Things (IoT) stimulates the need for timely status updates. For example, health and meteorological data need to be updated using the data collected by remote sensors in a timely manner~\cite{Chung2007,Liu2007}. To this end, the communication network should be designed to ensure timely status update, which can not be directly achieved through optimizing traditional metrics, such as outage probability, throughput, and queueing delay. Recently, a new metric, namely the \emph{Age of Information} (AoI)~\cite{Yates2019}, is proposed to evaluate the freshness of the received data. Existing studies on AoI focus on wireline networks with fruitful theoretical insights established based on queueing theories~\cite{Molta2020} and recent attention have been turned to wireless networks~\cite{Liu2022}. Clearly, timely delivery of status packets is challenging in wireless networks due to time-varying wireless channel conditions.

In many IoT applications, the information sink node might be distant from the sensing devices or their direct links may be blocked by large obstacles that make timely status updates difficult. To remedy, intermediate relays can be deployed to extend the effective communication range of the sensing devices with limited transmission power. Information relaying using cooperative relays has been shown to achieve spatial diversity without the need for a large antenna array~\cite{Blet2006}, which is not practical for sensing devices with small form factors. Some existing work has investigated how to enable status updates using relays. In~\cite{Mora2020}, the authors analyze the AoI of the wireless relay network based on the theoretical tool of Markovian jump linear systems. A static link scheduling policy, which assigns the transmission opportunity to either the source or the relay nodes with a fixed probability, is considered in their work. Since the transmission probability for AoI minimization does not have a closed form, it is optimized using numerical search in~\cite{Mora2020}. In~\cite{Xie2021-1, Xie2021-2}, conventional relay selection (RS) schemes are applied to the dual-hop status update network with a focus on the tradeoff between the average AoI and the average energy cost. In the aforementioned work, relays are assumed to have a permanent power supply, which requires frequent battery change or costly power cabling. For IoT use cases, an appealing implementation choice is to deploy EH relays that replenish energy from the source signal. \cite{Perera2020} explores the use of EH relay for real-time status updates in IoT by leveraging the notion of simultaneous wireless information and power transfer (SWIPT). However, their work is limited to a single-relay case and thus no spatial diversity is exploited for status updates.

In light of the explosive demand for real-time sensing in IoT, this work studies the AoI minimization problem in the dual-hop status update network that has not been well addressed. In this context, efficient status updates can benefit from selection diversity, namely choosing the best relay to receive and transmit the status update packet, respectively. Unlike~\cite{Xie2021-1, Xie2021-2} where the considered RS schemes are based on channel status information (CSI) without a specific design for ensuring information freshness, the RS problem for AoI minimization is solved in this work by a model-free approach based on deep reinforcement learning (DRL). The proposed DRL-based RS scheme makes the RS decision by jointly considering CSI packet freshness and buffer evolution relevant to the AoI. The advantage of the proposed DRL-based RS scheme is that it bypasses the difficulty of handling a large state space encountered in the Markov decision process (MDP) based formulation in order to arrive the optimal RS decision.

The remainder of this paper is organized as follows. Sec.~\ref{sec: system-model} illustrates the system model and problem formulation. The proposed DRL-based RS scheme is described in Sec.~\ref{sec-ddqnper}. Sec.~\ref{sec-results} presents the numerical results. Sec.~\ref{sec-conclusions} summarizes the key results. \textbf{\textit{Reproducible research}}: All the simulation results can be reproduced using Python code available at: 
https://github.com/Sinyu104/Relay-Selection-based-on-DQN.

\section{System Model and Problem Formulation}\label{sec: system-model}

We consider a dual-hop status update network where a source node $S$ sends status update packets to a destination node $D$ with the aid of $K$ intermediate relay nodes $R_1, \cdots, R_K$ using decode-and-forward (DF). A status packet is generated in a slot with probability $\lambda$. The slot length is chosen to allow at most one packet transmission. All nodes are single-antenna half-duplex devices. The direct communication between $S$ and $D$ is not available due to severe blockage. Therefore, $D$ can only receive data via relays that are self-sustainable by replenishing energy from the radio frequency (RF) signal sent from $S$. To this end, each relay has an energy-harvesting (EH) circuit that converts the RF signal into DC with a conversion efficiency $\eta \in (0, 1)$~\cite{Khan2020}. The received packet and energy is stored in a data buffer and an energy buffer, respectively, both with a finite storage~\cite{Lin2021}. The data buffer can accommodate at most one packet, because backlogged packets in the data buffer is not helpful to status update~\cite{Zake2021}. Each packet stored in $R_{k}$ consists of a time stamp $\alpha_{k}$ recording its generation time. The energy buffer has a size of $E_{\max}$ in the unit of energy intervals. Here, an energy interval is equal to the energy consumption for transmitting one packet by the relay. The instantaneous lengths of the data and the energy buffers of relay $R_k$ are denoted as $\delta_k\in\{0,1\}$ and $\epsilon_k\in\{0,1,\cdots,E_{\max}\}$, respectively. The constant power transmitted by the source and the relays is denoted as $P_S$ and $P_R$, respectively. 


The transmission over each link experiences path-loss and multi-path fading. Accordingly, the channel coefficients between $S-R_k$ and $R_k-D$ links, denoted as $h_k$ and $g_k$ for $k=1,\cdots, K$, are modeled as independent and identically distributed (i.i.d) zero-mean complex Gaussian random variables with variances $d^{-2}_{h_k}$ and $d^{-2}_{g_k}$ with $d_{h_k}$ and $d_{g_k}$ being the distance between $S$ and $R_k$ and that between $R_k$ and $D$, respectively. The instantaneous signal-to-noise ratio (SNR) of the $S-R_k$ link is symbolized as $\gamma_{h_k} \triangleq  | h_{k} |^2 \frac{P_S}{N_0}$ where $N_0$ denotes the noise power. Similarly, the instantaneous SNR of the $R_k-D$ link is represented as $\gamma_{g_k} \triangleq | g_{k} |^2 \frac{P_R}{N_0}$. Denote $t$ the slot length, corresponding to the time duration for transmitting one packet. The amount of energy harvested by $R_k$ in one slot is given by
\begin{align} \label{eq: harvested energy}
E_k 
=\eta P_{S}\left | h_{k} \right |^2 \cdot t.
\end{align}


\subsection{Transmission Protocol}\label{sec:trans-prot}
To leverage relays for status update, the following protocol is used. In each time slot, the DRL agent at $S$ decides whether to activate an $S-R$ link or an $R-D$ link. If an $S-R$ link is activated, $S$ transmits a packet to a selected relay and the remaining relays replenish energy from the source signal. The received packet by the relay will be stored in the data buffer before transmission. If an $R-D$ link is activated, a selected relay will forward a packet to $D$ while other relays remain idle.

In measuring information freshness at $D$, we use AoI as the metric~\cite{Yates2021}. For slot $n$, the AoI denoted as $A(n)$ is the time elapsed since the generation time of the last successfully received packet by $D$, i.e, $A(n)=n-\hat{\alpha}(n)$, where $\hat{\alpha}(n)$ is the generation moment of the latest received update packet by $D$ at slot $n$. The packet forwarded from a relay might be outdated, i.e., $\hat{\alpha}(n')<\hat{\alpha}(n)$ for $n'<n$ in which case AoI will be accumulated. Otherwise, AoI is reset to the time that elapsed since the generation of the latest packet. In summary, the AoI at slot $n$ is given by
\begin{align} \label{eq: age}
A(n)=\begin{cases}
n-\hat{\alpha}(n), &~\hat{\alpha}(n)>\hat{\alpha}(n'), n > n'\\
\min\{A(n-1)+1,A_{\max}\}, &~\text{otherwise}.
\end{cases}
\end{align}
where $A_{\max}$ is a prescribed upper bound. A packet received at $D$ with the AoI exceeding $A_{\max}$ will be discarded. Also, the packet in the relay buffer might be outdated when it is generated later than that of the latest received packet by $D$. To measure the freshness of the buffered packet at relay $R_k$, define the relative age as $\alpha'_k(n) \triangleq \alpha_{k}(n)-\hat{\alpha}(n)$~\cite{Zake2021}.


\subsection{Problem Formulation}\label{sec-problem}

The agent at $S$ aims to minimize the average AoI by selecting either a reception relay denoted as $R_r(n)$ or a transmitting relay denoted as $R_t(n)$ in slot $n$. For a time window of $N$ slots, the average AoI minimization problem can be formulated as follows.
\begin{subequations}
\begin{flalign}
(\text{P})  & \quad \mathop{\min} \limits_{\boldsymbol{\mathcal{R}}} ~\frac{1}{N} \sum_{n=1}^N A(n) &\nonumber \\
\text{s.t.}
&\quad  \delta_k(n) \in \{0,1\}, &\label{eq:data-limit}\\
&\quad 0 \leq \epsilon_k(n)\leq E_{\max}, &\label{eq:energy-limit}\\
&\quad  \delta_{R_{t}}(n) = 1~\text{and}~\epsilon_{R_{t}}(n) \geq 1,  &\label{eq:trans-limit}\\
&\quad  \gamma_{R_{r}}(n) \geq \gamma_{\text{th}}~\text{and}~\gamma_{R_{t}}(n) \geq \gamma_{\text{th}},  &\label{eq:snr-limit} \\
&\quad \alpha'_{R_t}(n)>0. &\label{eq:age-limit}
\end{flalign}
\end{subequations}
for $k=1,\cdots, K$ and $n=1,\cdots, N$. In (P), $\boldsymbol{\mathcal{R}}$ is an $N\times1$ vector with each entry indicating the index of the selected relay. Constraints (\ref{eq:data-limit}) and (\ref{eq:energy-limit}) follow due to the size limit of the data and energy buffers, respectively. Constraint (\ref{eq:trans-limit}) ensures that the transmitting relay has a non-empty data buffer and its available energy is sufficient for transmitting a packet. The activation of an $S-R$ link and an $R-D$ link takes place when their instantaneous SNRs exceeds a threshold $\gamma_{\text{th}}$, as stated in \eqref{eq:snr-limit}. Finally, constraints (\ref{eq:age-limit}) guarantees that an outdated packet would not be sent to $D$. Problem (P) can be treated as a MDP and solved by algorithms based on lookup tables (e.g., dynamic programming and reinforcement learning). However, the state space required to capture the channel and buffer status (see Sec.~\ref{sec-dqn-state}) can be extremely large that motivates us to develop a DRL-based approach.

\section{DDQN-PER Relay Selection Scheme}
\label{sec-ddqnper}

To minimize the average AoI, we propose a centralized model-free method based on deep Q network (DQN), which leverages neural networks in place of look-up tables to find $\boldsymbol{\mathcal{R}}$ while improving the algorithm convergence and stability through double DQN (DDQN) with priority experience replay (PER). We refer the proposed scheme as DDQN-PER.

\subsection{State Space}
\label{sec-dqn-state}

In making the RS decision, the agent at $S$ learns the environment from the following four pieces of information: the CSI of the $S-R$ links and the $R-D$ links, the data and the energy buffer status. Firstly, the instantaneous CSI is parameterized by the normalized channel gains as given by $\bar{\boldsymbol{\mathcal{H}}}(n) = \frac{\boldsymbol{\mathcal{H}}(n)}{\max\;\boldsymbol{\mathcal{H}(n)}}$ where $\boldsymbol{\mathcal{H}}(n)= [|h_1(n)|~\cdots~|h_{R_K}(n)|]^T$ for the $S-R$ links and $\bar{\boldsymbol{\mathcal{G}}}(n) = \frac{\boldsymbol{\mathcal{G}}(n)}{\max\;\boldsymbol{\mathcal{G}(n)}}$ where $\boldsymbol{\mathcal{G}}(n)= [|g_1(n)|~\cdots~|g_{R_K}(n)|]^T$ for the $R-D$ links. Provided with CSI, the agent can make a decision to meet constraint \eqref{eq:snr-limit} while maximizing the total amount of harvested energy by relays. For example, if two relays satisfy constraint \eqref{eq:snr-limit}, the one with a higher channel gain of the $S-R$ link could charge more energy and thus it can be selected to replenish energy while the other relay is selected to receive a packet from $S$. Clearly, the amount of accumulated energy depends on the available room of the energy buffer. For slot $n$, the energy buffer status is represented by $\bar{\boldsymbol{\mathcal{E}}}(n) = \frac{\boldsymbol{\mathcal{E}}(n)}{E_{\max}}$ where $\boldsymbol{\mathcal{E}}(n)=[\epsilon_1(n)~\cdots~\epsilon_{R_K}(n)]^T$. Knowing $\bar{\boldsymbol{\mathcal{E}}}(n)$ is also necessary to meet the energy buffer constraint in~\eqref{eq:trans-limit}. Similarly, the data buffer constraint in~\eqref{eq:trans-limit} is handled by feeding the data buffer status denoted as $\boldsymbol{\mathcal{D}}(n)=[\delta_1(n)~\cdots~\delta_{R_K}(n)]^T$ to the agent. When more than one relays satisfy the constraints (\ref{eq:trans-limit}) and (\ref{eq:snr-limit}), the age information is essential for decision-making. Therefore, we include the normalized relative age denoted as $\bar{\boldsymbol{\mathcal{A}}}(n) = \frac{\boldsymbol{\alpha'}(n)}{A_{\max}}$ in the state, where ${\boldsymbol{\alpha}}'(n)=[\alpha'_1(n), \cdots, \alpha'_{K}(n)]^T$ is the vector of relative age. In addition, a binary indicator denoted as $I(n)$ is employed to indicate whether there is a packet arrives in slot $n$, which takes the value of one when there is a new packet generated in slot $n$ and it is zero otherwise. Overall, the state of slot $n$ is represented as $s(n) \triangleq \{\bar{\boldsymbol{\mathcal{H}}}(n), \bar{\boldsymbol{\mathcal{G}}}(n), \boldsymbol{\mathcal{D}}(n), \bar{\boldsymbol{\mathcal{E}}}(n), \bar{\boldsymbol{\mathcal{A}}}(n),I(n)\}$. While $\boldsymbol{\mathcal{D}}(n)$, $\bar{\boldsymbol{\mathcal{E}}}(n)$ and $\bar{\boldsymbol{\mathcal{A}}}(n)$ are finite in dimensions, $\bar{\boldsymbol{\mathcal{H}}}(n)$ and $ \bar{\boldsymbol{\mathcal{G}}}(n)$ have infinitely large dynamic ranges. Consequently, the dimension of $s(n)$ is infinite. 

\subsection{Action Space}

At the beginning of each slot, the agent picks an action to determine which of the $2N$ links is activated. Let $R^*(n)$ denote the relay associated with the selected link in slot $n$, where $R^*(n)$ is a transmitting relay if an $R-D$ link is activated or a receiving relay when an $S-R$ link is activated, whichever maximizes the reward as will be specified later. Accordingly, the action of slot $n$ can be expressed as $a(n)=R^*(n)$ and the action space of $N$ slots is given by $\mathcal{A}=\{a(1),\cdots,a(N)\}$.

\subsection{Reward Function}

Based on $s(n)$, the agent takes a specific action and receives a certain reward according to a reward function. To minimize the AoI, it is better to maximize packet freshness at the relay side. Noted that based on the definition of relative age in Sec.~\ref{sec:trans-prot}, the positive value of relative age means the packet in the relay buffer is fresher than the packet received at $D$ and the negative value of relative age means the packet is outdated. Thus, the action of activating the $S-R$ link is assigned with the reward equal to the accumulated relative age of all the packets in the relay buffers, as given by
\begin{align}\label{eq:dqn-action-recep}
r_a(n)= \sum_{k=1}^{K} \alpha_{k}^{'}(n).
\end{align}
As for the $R-D$ link, the action of transmitting the freshest packet should be encouraged. This can be achieved by a reward function given as
\begin{align}\label{eq:dqn-action}
r_a(n)= \alpha'_{R_t}(n).
\end{align}
We note that some actions may be infeasible, when (i) any of the constraints \eqref{eq:energy-limit}-\eqref{eq:age-limit} are violated, (ii) there is a packet generated at $S$ but the action of activating the $R-D$ link is chosen, and (iii) there is no packet generated at $S$ but the action of activating the $S-R$ link is chosen. To remedy, an infinitely small negative reward is assigned to the infeasible action.


\subsection{Proposed DDQN-PER-based RS Algorithm}

In the proposed DDQN-PER-based RS algorithm, the agent employs two convolutional neural networks (CNNs), namely the main network $Q$ and the target network $\hat{Q}$~\cite{van2015}, for determining the predicted Q-value and the target Q-value, respectively. These two networks share the same structure but with different parameters. The main network updates its parameters denoted as $\theta$ in each step while the target network synchronizes its parameters denoted as $\theta^-$ with $\theta$ of the main network periodically. In the training procedure, the agent executes the action determined by the maximum Q-value of the main network that is updated in every step to explore the unknown environment. Specifically, the system state changes from the current state $s(n)$ to the next state $s(n+1)$ when the agent selects an action $a(n)$ and earns an reward $r(n)$. The target Q-value in slot $n$ is 
\begin{align} \label{eq:ddqn-target}
y(n) = r(n+1)+\omega~[\hat{Q}(s(n+1), \arg\max_{a(n)}Q(s(n+1), a(n))]
\end{align}
where $\omega \in [0,1]$ is the discount factor and $Q(\cdot)$ and $\hat{Q}(\cdot)$ represent the action-value function of main network and target network, respectively. The mean square error between the target value and the predicted value of all transitions is defined as the loss function given by 
\begin{align}\label{eq:ddqn-loss}
L(\theta) = E \bigl[ y(n)-Q\bigl(s(n), a(n)\bigr) \bigr]^2.
\end{align}
By performing gradient descent to $L(\theta)$, the main network updates its parameters in each step. 


\subsection{Prioritized Experience Replay}
\label{sec-per}
Instead of sampling the experience uniformly, more efficient learning can be achieved by sampling the experience according to its importance, which is measured by the magnitude of the loss function of each transition. Let $L_i(\theta)$ denote the loss function of transition $i$, which is assigned with priority $p_i = |L_i(\theta)|+ \zeta$, with $\zeta$ being a small positive constant to avoid the transition with a zero loss function value never getting sampled. Given $p_i$, the probability of transition $i$ is
\begin{align}\label{eq:per-trans-prob}
P_i=\frac{p_i^\alpha}{\sum_j p_j^\alpha} 
\end{align}
where $\alpha\geq 0$ is used to control the weight of priority in the sampling process. The case $\alpha=0$ corresponds to uniform sampling. Clearly, prioritized replay changes the sample distribution that leads to biased estimate of the solution. The bias can be fixed by applying importance-sampling weights as given by~\cite{Schaul2015} 
\begin{align}\label{eq:per-isweight}
w_i = (\frac{1}{\mathcal{RM}}\cdot \frac{1}{P_i})^{\beta}
\end{align}
where $\mathcal{RM}$ is the replay memory size and $\beta$ is an exponent that controls how much prioritization to apply. The case $\beta=1$ reduces to uniform sampling without prioritization.

The complete DDQN-PER RS algorithm is presented in Algorithm.~\ref{alg:ddqn-per}. In Line~\ref{ddqn:line:make-decision1}-\ref{ddqn:line:store}, action $a(n)$ is selected based on $\varepsilon$-greedy and the corresponding experience is stored in the replay memory. Line~\ref{ddqn:line:transition1}-\ref{ddqn:line:transition2} calculate the loss function, the importance-sampling weight, and transition probability for each transition in the minibatch and accumulate the loss function value with weight $\theta$. In Line~\ref{ddqn:line:update}, the main network is updated. Lastly, the target network is periodically updated every $C$ steps. 

\begin{algorithm}
\caption{DDQN-PER-based relay selection algorithm}\label{alg:ddqn-per}
\begin{algorithmic}[1]
\STATE \textbf{Input:} mini-batch size $\mathcal{M}$, step-size $\mu$, replay memory size $\mathcal{RM}$, exponents $\alpha$ and $\beta$.
\STATE Initialize replay memory, $\Delta=0$, $p_1=1$

\FOR{$\text{Episode}=1$ to $E$}
\STATE Initialize each data buffer and energy buffer.
\FOR{$n=1$ to $N$}
\STATE Select $a(n)=\arg \max_{a}\mathcal{Q} \allowbreak (s(n),a)$ with probability $\varepsilon$ or a random action with probability $1-\varepsilon$. \label{ddqn:line:make-decision1} 
\STATE Execute action $a(n)$ and observe reward $r(n)$ and the corresponding state $s(n+1)$\label{ddqn:line:make-decision2}
\STATE Store $z_n = \{s(n), a(n), r(n),s(n+1)\}$ with maximal priority $p_n=\max_{j<n} p_j$ in replay memory \label{ddqn:line:store} 
\FOR{$i$=1 to $\mathcal{M}$}\label{ddqn:line:transition1}
\STATE Sample transition $i$ by probability $P_i$ in~\eqref{eq:per-trans-prob}.
\STATE Compute $w_i$ using~\eqref{eq:per-isweight}.
\STATE Compute loss function value $L_i(\theta)$ in~\eqref{eq:ddqn-loss}
\STATE Update transition priority $p_i \leftarrow |\rho_i|+ \zeta$
\STATE Accumulate weight change $\Delta \leftarrow \Delta+w_i L_i(\theta) \Delta_{\theta} \times Q(s_i, a_i)$ 
\ENDFOR\label{ddqn:line:transition2}
\STATE Update weights $\theta \leftarrow \theta + \mu \cdot \Delta$, reset $\Delta=0$\label{ddqn:line:update}
\STATE Reset $\theta^{-}=\theta$ every $C$ steps
\ENDFOR
\ENDFOR
\end{algorithmic}
\end{algorithm}

\section{Results and Discussions}
\label{sec-results}

\subsection{Simulation Setting}
\label{sec-simu-set}

In implementing DDQN, the main network $Q$ and the target network $\hat{Q}$ have the same structure comprising of an input layer with dimension $K \times 4$, a fully connected layer with eighty nodes using a rectified linear unit (ReLU) activation function, and an output layer with the dimension of $|\mathcal{A}|\times 1$. 
The DQN is trained using 1,000 episodes with 2,000 steps in each episode. The discount factor $\omega$ is set to one, the prioritized factors $\alpha=0.6$ and $\beta=0.4$. The size of the replay memory $\mathcal{RM}$ and mini-batch $\mathcal{M}$ is 10,000 and 32, respectively. The term ``SNR'' in the figures is defined as $\text{SNR}=P_{S}/N_0$, which is set to 70 dB, the noise power $N_0=0.01$, and transmission power ratio $P_S/P_R=1,000$. For correct decoding, the SNR threshold is equal to $2^{2R-1}$ where $R=1$ bit/sec/Hz is the target transmission rate. The packet arrival probability is $\lambda=0.3$ and $A_{\max}=100$. $E_{\max}=3$ and the RF-DC conversion efficiency $\eta=0.5$ for all the relays. Initially, each data buffer is empty, and each energy buffer is filled with one energy interval. All the links have the same distance of 36 m.



\subsection{Benchmarks}
\label{sec-bench}
We compare the performance of the proposed DQN-PER RS scheme with three benchmark schemes. 
\begin{itemize}
\item \textit{Dual Buffer Relay Selection} (DBRS)~\cite{Lin2021}, which is a heuristic scheme for minimizing the outage probability of buffer-aided relay networks. This scheme is chosen to verify whether minimizing the outage probability is beneficial to AoI minimization and it is similar to the proposed scheme with both CSI and buffer status considered.
\item \textit{Greedy} scheme, which makes the RS decision based on the generation time of each packet. The relay that has the latest generation time is selected to transmit. As to the reception relay, the relay with an empty data buffer has the highest priority to receive a status packet from $S$. If more than one relay has an empty buffer, the tight is broken by a random selection. If more than one relay has a non-empty buffer, the relay with the earliest generated packet is selected as $R_r$. 
\item \textit{Max-Link} scheme~\cite{Krikidis2012}, which activates a link that has the highest SNR among all eligible $S-R$ and $R-D$ links, where an eligible link must satisfy both constraints (\ref{eq:trans-limit}) and (\ref{eq:snr-limit}). This selection rule is similar to the proposed scheme but none of the buffer status is considered.
\end{itemize}
We note that \textit{DBRS} and \textit{Greedy} activate the $S-R$ link and the $R-D$ link in a fixed order while the order can be arbitrary in \textit{Max-Link} and the proposed schemes. The activation order would affect the status update performance as will be shown later. Also, if the selected relay by \textit{DBRS} and \textit{Greedy} schemes violates constraints (\ref{eq:trans-limit}) or (\ref{eq:snr-limit}), a single relay is chosen as: $R^*= \max_{\epsilon_{R_k}\neq 0, \delta_{R_k} < D_{\max} }~\min\{\gamma_{h_{k}}, \gamma_{g_{k}}\}$~\cite{Lin2021}. 

\subsection{Performance evaluation}
\label{sec-perf-com}

\begin{figure}[!t]

\centering
{
\includegraphics[width=0.7\linewidth]{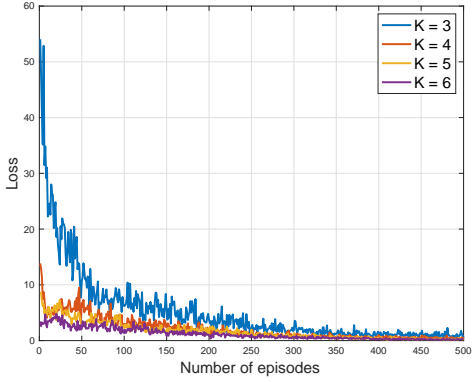}
}
\caption{Training loss of the proposed DDQN-PER model.}
\label{fig:cong}
\end{figure}

Fig.~\ref{fig:cong} shows the training result of the proposed DDQN-PER model by plotting the loss function values for 500 episodes. One can clearly observe that the proposed DRL algorithm converges for varied number of relays $K$. For a greater $K$, the loss function curve tends to converge ealrier.

\begin{figure}[!t]
\centering
\includegraphics[width=0.7\linewidth]{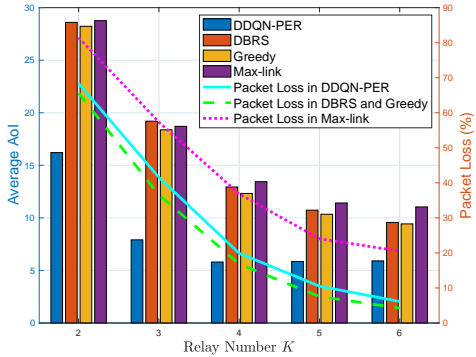}
\caption{Average AoI and packet loss rate \textit{vs.} number of relays $K$.}
\label{fig:diffrelaynum}
\vspace{-0.5cm}
\end{figure}

Fig.~\ref{fig:diffrelaynum} depicts the average AoI of different RS schemes versus the number of relays $K$. As $K$ increases, the average AoI of all the RS schemes is effectively reduced, thanks to the increased selection diversity. It is noticed that the proposed scheme has a higher packet lost rate (the ratio between the number of discarded packets by relays and the total number of packets generated by $S$) than \emph{DBRS} and \emph{Greedy} schemes but its average AoI is significantly lower. However, the same gain is not attained using \emph{Max-Link} scheme, which has a higher packet loss rate. This suggests that properly dropping packets helps to improve the AoI performance. For all the schemes, the gain from increasing $K$ is minor. This can be explained because the status packet is backlogged at relays when none of the relays has a sufficient amount of energy to transmit and this chance does not diminish as $K$ increases~\cite{Lin2021}. It is worth to point out \emph{Max-Link} scheme performs the worst when $K$ is large since it chooses to activate a link without considering packet freshness. \emph{DBRS} also does not consider packet freshness but it performs better than \emph{Max-Link} by further considering buffer status. Although \emph{Greedy} scheme considers packet freshness, it lacks of considering channel and buffer status that might end in choosing an infeasible action.

\begin{figure}[!t]
\centering
\vspace{-0.5cm}
\includegraphics[width=0.7\linewidth]{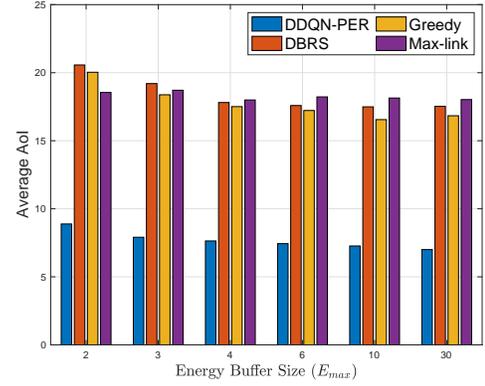}
\caption{Average AoI \textit{vs.} energy buffer size $E_{\max}$ for $K=3$ relays.}
\vspace{-0.5cm}
\label{fig:diffEngsize}
\end{figure}

The impact of energy buffer size $E_{\max}$ is investigated in Fig.~\ref{fig:diffEngsize} for $K=3$ relays. Intuitively, a larger energy buffer helps to accumulate more energy thereby less likely that none of relays can transmit. From the figure, only marginal reduction of the average AoI is achieved by increasing the energy buffer size since the amount of harvested energy is relatively small compared to that required for transmitting a packet. In other words, the energy consumption rate is much higher than the supply rate and this unbalance can not be mitigated by a large energy buffer size.

\section{Conclusion}\label{sec-conclusions}
This work addressed the sophisticated RS problem for AoI minimization in the EH relay-assisted status update network by proposing a DQN based approach. Simulation results demonstrated that the proposed RS scheme effectively improves the average AoI than the other competitive schemes. When the number of relays is small, the proposed scheme reduces about $50\%$ average AoI of existing RS schemes. Among numerous factors that affect the AoI performance, adding many relays and the size of energy buffers only has limited improvement to the average AoI.

\bibliographystyle{IEEEtran}

\end{document}